\documentclass[epjST]{svjour}
\usepackage{doi}
\usepackage{hyperref}
\hypersetup{
 colorlinks=true,
 linkcolor=blue,
 filecolor=magenta, 
 urlcolor=cyan,
 pdftitle={Molecular 9Be-p Fusion Reaction},
 }
\usepackage{graphicx}
\usepackage{amsmath,amssymb,amsfonts}
\newcommand{\be}{\begin{equation}}
\newcommand{\beq}{\begin{equation}}
\newcommand{\ee}{\end{equation}}
\newcommand{\eeq}{\end{equation}}
\newcommand{\ba}{\begin{eqnarray}}
\newcommand{\ea}{\end{eqnarray}}
\newcommand{\ban}{\begin{eqnarray*}}
\newcommand{\ean}{\end{eqnarray*}}
\newcommand \nn {\nonumber}
\newcommand{\req}[1]{Eq.\,({\ref{#1}})}
\newcommand{\rf}[1]{Fig.\,{\ref{#1}}}
\newcommand{\orcA}{0000-0001-8217-1484}
\newcommand{\orcB}{0000-0003-2468-3996}
\newcommand{\orcidicon}{\includegraphics[width=0.32cm]{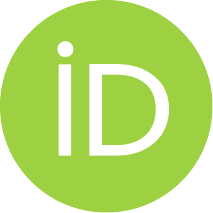}}
\newcommand{\orc}[1]{\href{https://orcid.org/#1}{\orcidicon}}

\begin{document}
\author{
Johann Rafelski${}^1$\orc{\orcA}\thanks{Corresponding author 
\email{johannr@arizona.edu}}, 
Berndt M\"uller${}^{2}$\orc{\orcB}
}
\institute{${}^1$Department of Physics, The University of Arizona, Tucson, AZ, 85721, USA\\
${}^2$Department of Physics, Duke University, Durham, NC, 27708, USA}

\title{Molecular $^9\mathrm{Be} + p$ Fusion Reaction} 

\date{\today}

\abstract{We study (nuclear) fusion reactions in Beryllium. We argue that some are nuclear long-distance molecular processes of interest for low energy nuclear reactions. For these we develop a novel reaction model. We note a chain of reactions that can naturally arise in Beryllium target created by nonequilibrium proton reactions.}

\maketitle
\section{Introduction and Motivation}
 Nuclear aneutronic `fusion' reactions with only charged particles in the final state are of primary interest for civilian fusion energy production. In these reactions either a bound neutron (typically attached to a proton in the incoming deuteron) or just a proton are inducing a reaction capable of releasing nuclear energy $Q$. For the neutron transfer the most often considered such aneutronic reaction is 
\be
^3\mathrm{He}+d\to\! ^4\mathrm{He}+p + Q \,,\qquad Q=18.35304\,\mathrm{MeV}\,.
\label{eq:dhe}
\ee
We use nuclear data seen at Nuclear Data Evaluation Project st TUNL (Triangle Universities Nuclear Laboratory) \url{https://nucldata.tunl.duke.edu/nucldata/index.shtml}. Similarly, the absorption of a proton on Boron 
\be
^{11}\mathrm{B}+p\to\! ^4\mathrm{He}+ ^{8\!\!}\mathrm{Be} + Q \,,\qquad Q= 8.5903 \,\mathrm{MeV}\,.
\label{eq:pBe}
\ee
is a widely studied aneutronic reaction. The theoretical understanding of these reactions has been largely achieved in the framework of the R-matrix theory many years ago and relevant details are textbook material. 

We explore here yet another type of aneutronic reaction which we call molecular fusion
\be
^9\mathrm{Be}+p\to\! ^8\mathrm{Be}+d + Q \,,\qquad Q=0.5592\,\mathrm{MeV}\,.
\label{eq:9Bep-8Ben}
\ee
where the neutron transfer proceeds in the `opposite' direction: An incoming slowly moving proton $p$ picks up a fast molecular neutron in a collision reaction process. The nucleus $^9\mathrm{Be}$ can be seen to consist of two $\alpha$-particles bound by a `molecular' $J=3/2^-$ neutron $n$. We will refer to such loosely  bound neutrons in open shells as `valence neutrons', in analogy to the use of the term `valence electrons' in atomic physics, as they can facilitate the binding of relatively inert nucleon clusters, such as $\alpha$-particles. In this reaction a newly fused deuteron $d$ emerges. The reaction~\req{eq:9Bep-8Ben} is the only ``inverse'' exothermic transfer reaction in light nuclei that can produce significant reaction energy having $Q>0$. 

However, many slightly endothermic reactions exist which could be a stepping stone within a cycle leading to civilian fusion energy production. Therefore our present work opens a novel methodology which can relatively easily be generalized to other reaction cycles. For energy gain to occur in reaction~\req{eq:9Bep-8Ben} the valence neutron in the heavier nuclear target $^9\mathrm{Be}$ has to be less strongly bound with separation energy $E_s^{(n^8\mathrm{Be})}=-1.6654$\,MeV compared to the produced $d$, where it is already relatively weakly bound $E_d^{(np)}=-2.22457$\,MeV; in transfer of the charge neutral neutron $n$ the net exothermic energy gain ($Q$-value) shown in~\req{eq:9Bep-8Ben} is created. 

Reaction~\req{eq:9Bep-8Ben} has been for a long time recognized as needing additional theoretical attention due to not fully understood low $p$-energy reactivity as characterized by the large value of the Coulomb suppression factor (S -factor). Considering the nuclear-molecular structure of $^9$Be ~\req{eq:9Bep-8Ben} resembles a molecular reaction process. We can use some of pertinent methods and for this reason we call the neutron transfer reaction molecular fusion: A neutron transfer occurs over a considerable distance, which implies that pivotal features could remain ignored in the R-matrix approach. We will proceed in this study assuming that the nuclear reaction~\req{eq:9Bep-8Ben} is more akin to a chemical processes with the neutron transferring to proton at a relatively large distance, invalidating the short range R-matrix nuclear process hypothesis. 

This weak binding makes the static wave function of the bound neutron have a long-range tail provoking questions about the validity of the R-matrix method: The R-matrix method was invented to describe nuclear transformation occurring at a well defined nearly sharp nuclear surface. 
 
What invalidates for the reaction~\req{eq:9Bep-8Ben} the R-matrix model are two circumstances:
i) A smooth and slow transition between the asymptotic in-out reaction states akin to situation known from molecular reactions as will be shown in detail. This is possible since the weak binding of the valence $n$ in $^9\mathrm{Be}=\alpha$-$\alpha$-$n$ makes the static wave function of the bound neutron have a long-range tail.
ii) Unlike the case of atomic reactions we will recognize a more shallow and relatively longer ranged dynamic barrier: as the neutral but strongly interacting neutron `moves' from location near to $\alpha$-$\alpha$ where it forms $^9\mathrm{Be}$ to a better location around $\alpha$-$\alpha$-$p$ system emerging ultimately transferred to the fly-by $p$, there is a `price to pay' for a transient less optimal wave function shape even if and when in the asymptotic state $d=p$-$n$ the binding nearly doubles. The reader may now wonder with us if other neutron transfer reactions at a low collision energy of two colliding (a.k.a fusing) nuclei need to be reconsidered in a molecular reaction model of the kind we develop in this work. Indeed, it seems that similar situation arises whenever the relative collision speed is slower when compared to the intrinsic motion of the transferring neutron. We note that this is {\em opposite} to reactions involving `fast' collisions where similar language and theoretical methods appear~\cite{Moschini:2021}.

Another aneutronic fusion reaction involving $^9\mathrm{Be}$ is (for 2012 update to produced reaction energy see Ref.\,\cite{Kelley:2012}) 
\be
^9\mathrm{Be}+d\to\! ^{10}\mathrm{Be}+p + Q \,,\qquad Q= 4.5877\,\mathrm{MeV}\,.
\label{eq:9Bed-10Bep}
\ee
The produced nucleus is a molecular state of two $\alpha$-particles glued together by two valence neutrons which are both found in the same molecular $J=3/2^-$ orbit as was the case in $^9\mathrm{Be}$. The required $0^+$ symmetric spin state is possible since the ground state and several excited states of $ ^{10}\mathrm{Be}$ have symmetric isospin $T= 1$, the antisymmetric character of the two valence neutrons is carried by the spin-orbital part of the wave function. This reaction~\req{eq:9Bed-10Bep} is of interest to us as it could be induced by a $d$ produced in primary reaction~\req{eq:9Bep-8Ben}.

There is indeed a chain of reactions which could be of interest in the context of laser pulse driven nonequilibrium fusion energy production involving $^9\mathrm{Be}$: In a first step energetic $d$ are produced in reaction~\req{eq:9Bep-8Ben}. These deuterons carry away as seen in laboratory both the incoming proton $p$ energy, and most of the energy produced. This allows these produced $d$ to induce in resonant manner the sequel reaction~\req{eq:9Bed-10Bep}. The produced $p$ are at an energy which allows multitude of tertiary reactions needing further study. The other secondary reaction product $^{10}\mathrm{Be}$ $\beta$-decays with half life of 1.5 million years into $^{10}\mathrm{B}$ releasing observable $Q_\beta=0.556$\,MeV in form of $\beta^-$-radiation. Given the long life-span accumulation of $^{10}\mathrm{Be}$ in a fusion target can be expected. In fact for this very reason $^{10}\mathrm{Be}$ is used as a time marker in geology; we note a recent review and abundance anomaly report~\cite{Koll:2025} indicating need for further exploration of the reactions we explore here in order to ascertain all origins of this important isotope. 

We recall that we have in a fusion target both `beam' of protons  $p$  (e.g. laser pulse generated) and protons produced in reaction~\req{eq:9Bed-10Bep} which can now induce on the produced $^{10}\mathrm{Be}$ another cycle closing reaction, the `pickup' of the di-neutron from $^{10}\mathrm{B}$ to form tritium $t$ carrying away most of the energy of the incoming proton
\be
 ^{10}\mathrm{Be}+p \to ^{8\!}\mathrm{Be}+t+ Q \,,\qquad Q= 0.0050\,\mathrm{MeV}\,.
\label{eq:9Bep-8Bet}
\ee
The reaction Q-value is computed from other reactions and is seen in Fig.\,4 in Ref.\,\cite{Kelley:2012}, its value is exceedingly small. Seen how degeneracy in energy for this di-neutron transfer process we will return to look more closely at this reaction under a separate cover. What is of interest is that much of the proton energy is kept in the produced $t$ which is a very active ingredient of many nuclear reactions and is hard to make and use as a nonequilibrium beam of non-thermal particles. It seems that the reaction~\req{eq:9Bep-8Bet} offers a no-cost path to create beams of $t$, capable to undergo additional reactions to be considered under separate cover. We further note that of course one step process combining reactions~\req{eq:9Bed-10Bep} and~\req{eq:9Bep-8Bet} can occur$^9$Be$+d\to ^8$Be$+t +4.592$\,MeV but this step is out of resonance at the available reaction energy.

In the following we introduce a few methods allowing a novel theoretical description of (di)neutron transfer reactions between light elements in relatives slow motion when compared to the (di)neutron and show some rudimentary results supporting the ideas described above.

\section{Molecular reaction model}
\label{sec:molecular}
\subsection{Tunneling instantons}
\label{ssec:tunnel}

The electrically neutral neutron transferring from $^9$Be to the incident proton encounters
the tunneling barrier consisting of the separation energy from the $\alpha$-$\alpha$ core while traveling into a deeper binding well surrounding the proton. The dynamical width of this barrier is controlled by the distance between the moving proton and $^9$Be. The height of this barrier corresponds to the separation energy of the neutron from the $\alpha$-$\alpha$ core, $E_s=1.6654$\,MeV. This is the key factor favoring $^9$Be$+p$ reactions, as among the light fusing nuclei neutron the binding to the $^8$Be=$\alpha$-$\alpha$ core is the smallest. The large size of the neutron orbit in $^9$Be reduces the width of the tunneling barrier at a given distance of the proton as compared to other fusion systems, and this is certainly true for the inverse transfer reaction~\req{eq:dhe} where the neutron needs to penetrate a higher barrier due to stronger neutron binding in deuterium. 

A neutron-transfer reaction involves tunneling across a dynamically changing barrier from one binding potential well to a more strongly binding potential well. In association with this process the reaction energy gain needs to be shared, in the case or reaction~\req{eq:9Bep-8Ben} between $^8$Be (2/10-fraction) and outgoing $d$ (8/10-fraction) to account for momentum conservation: In the nonrelativistic limit and the common center-of-mass system the momenta of the two particles traveling in opposite direction satisfy $m_1\vec v_1=-m_2\vec v_2$. This relation implies that the kinetic energies $E_i$ of the two reaction products $^8$Be and $d$ are related as $E_1=(m_2/m_1)E_2$ while energy conservation assures that $Q=E_1+E_2$. We thus find $E_2=Q m_1/(m_1+m_2)$ and similarly for the other fusion product. We remember that the lighter reaction product acquires a larger fraction of the released energy in any nuclear two-body reaction. 

Let us look at the exponential suppression that governs tunneling: on the $^8$Be-$n$ side the imaginary momentum is $|\tilde{P}_\mathrm{Be}|=\sqrt{2\mu_n |E_s|}= 52.7$\, MeV (here and in the following the symbol $\mu$ refers to reduced mass). On the deuteron ($p-n$) side we have $|\tilde{P_d}|=\sqrt{2\mu_n |E_d|}=45.7$\,MeV. The difference between these values relates to the four-body dynamics of the $\alpha$-$\alpha$-$n$-$p$ system. Since the difference in tunneling momentum is only $\pm6$\%, a complete treatment that accounts for this difference is not necessary in a first study. Therefore, in Section~\ref{ssec:transfer} we use methods developed to describe the transfer of (light) electrons in atomic collisions in order to construct a model of the dynamical barrier for neutron transfer. 

This conceptual analogy arises since the incoming proton will move slowly near to the classical turning point, its asymptotic kinetic energy being used to climb the Coulomb barrier. The virtual molecular motion of the neutron in $\alpha$-$\alpha$-$n$ under the tunneling barrier is fast compared with the $^9$Be-$p$ motion when the asymptotic proton energy is less than $|E_s|$, allowing for the molecular Born-Oppenheimer approximation with the ``fast'' neutron shared between $^8$Be and the incoming $p$. The usual molecular Born-Oppenheimer approximation preserves angular momentum conservation which assures that the transfer path involves a one-dimensional radial molecular coordinate. The interesting feature here is that the neutron tunneling between the two binding centers is effectively one-dimensional.

In such situations tunneling resonances, called instantons, arise that are controlled by the variable tunneling distance $x_t$. These instanton solutions are stationary states found in an inverted potential well. They can be thought of as bound states that arise when the Schr\"odinger equation is considered in imaginary time $\tau=it$. The width of the inverted potential determines the instanton energy and depends on the distance of the proton from the $^8$Be core and thus on the collision energy. At this resonant collision energy no attenuation of the neutron transfer probability arises in the tunneling process. The instanton enabled tunneling process thus can lead to a large reaction strength (see {\it e.g.\/}~\cite{Le_Deunff_2010}). 

The tunneling neutron wave function in the well domain can be described by a single instanton, an example is $\psi=N\cosh\{(x-x_t/2)\tilde P \}$, where $x$ is the coordinate of the neutron centered about midpoint $x_t/2$ in distance between the proton and the core $\alpha$-$\alpha$. The instanton momentum $\tilde P $ varies between the two instanton edges as discussed above $45.7\,\mathrm{MeV}=\tilde P_d<\tilde P<\tilde P_\mathrm{Be}=52.7\,\mathrm{MeV}$ due to the recoil effects. Fortunately in the system we are considering with $P_d\simeq P_\mathrm{Be}$ the influence of the exact many body dynamics is minimal since the real momentum of the neutron bound to $\alpha$-$\alpha$ core is nearly the same as in $d$. To add another surprise, the recoil momentum between the newly formed deuteron and remnant $^8$Be is also near this value, $P^\mathrm{recoil}=41\,\mathrm{MeV}$. In conclusion of this discussion we note a known resonance with $E_i=-25.9\pm1.9$\,keV with reference to, and below, the $p$-$^9$Be energy threshold ($6 560 \pm 1.9$\,keV above compound $^{10}$B ground state) with a width $\Gamma_i=25.1 \pm 1.1$\,keV . This could be just the here described reaction instanton. 

We conclude this discussion noting that the reaction~\req{eq:9Bep-8Ben} appears to harbor enhanced fusion reactivity for low energy collisions, even when compared to~\req{eq:dhe}. This suggests need for considerable further theoretical study of the low energy nuclear reaction. We make a step in this direction and now formulate the neutron transfer reaction in the molecular model using the semiclassical Born-Oppenheimer approximation used in the traditional meaning:  the nuclear motion is fast, while the motion of the binding particle, here the neutron, is fast.

\subsection{Molecular transfer}
\label{ssec:transfer}

The $^8{\rm Be}+d$ system should act as a $J^\pi = 1^+$ sub-threshold doorway resonance in $^9{\rm Be}+p$ scattering, which is visible in the partial wave analysis of the $^9$Be(p,d)$^8$Be reaction (see \cite{Barker:2001uvh}, Fig.~1b) in the rapid rise of the spectroscopic factor~\cite{Macfarlane:1960zz}, which is defined as the overlap between nuclear wave functions in the initial and exit channel, near threshold.

In order to study this reaction in more detail we consider a simple model based on the approach of Rakityansky \cite{Rakityansky:2024zew}, who modeled the $^9$Be ground state as a $\alpha\alpha$n cluster. In this model the rms separation of the two $\alpha$ clusters is found as $R_{\alpha\alpha} = 3.46$ fm, and the rms distance between the neutron and the center of mass is obtained as $R_{\rm n,CM} = 4.65$ fm. Our (extremely crude) model assumes that the neutron in $^9$Be and in the deuteron is bound by a spherical square potential. We fix the parameters of the potential to match the mean square radius of a pion Yukawa potential around a core with the electric charge radii of $^9$Be ($r_c = 2.52$ fm \cite{Nortershauser:2008vp}) and of the proton ($r_c = 0.84$ fm), respectively, and to yield bound states with the experimentally measured neutron binding energies, which are listed above. The square well radii are taken to match the mean square radii of the nuclear core, given by $r_c$, and mean square range of the Yukawa potential $V(r) = V_0 e^{-mr}/r$:
\be
\langle r^2 \rangle_{\rm sw}
= \frac{\int_0^R r^2 dr r^2}{\int_0^R r^2 dr } 
= \frac{3}{5}R^2,
\qquad\qquad
\langle r^2 \rangle_{\rm Y}
= \frac{\int_0^\infty r^2 dr\, r^2 V(r)}{\int_0^\infty r^2 dr\, V(r)} 
= \frac{\int_0^\infty r^3 dr\, e^{-mr}}{\int_0^\infty r dr\, e^{-mr}}
=\frac{6}{m^2}\,.
\ee
In the spirit of the meson exchange model of nuclear interactions \cite{Machleidt:1987hj} we choose the square well radii to reproduce the effective mean square radii $r_c$ of the $^9$Be core and the proton folded with a Yukawa potential, respectively. We further assume that the $p-n$ potential is dominated by one-pion exchange ($m = m_\pi$), whereas the range of the $^8{\rm Be}-n$ potential is given by scalar meson exchange with $m = m_\sigma \approx 600$ MeV \cite{Machleidt:1987hj}.
The resulting radii and potential depths are $R_{({\rm Be})} = 3.42$ fm and $V_0^{({\rm Be})} = -23.04$ MeV for $^9$Be, and $R_{d} = 4.65$ fm and $V_0^{(d)} = -10.31$ MeV for the deuteron.

The deuteron is a spin-1 triplet, with the dominant configuration being a $L=0$ bound state, while the valence neutron in $^9$Be is in a $J=3/2$, $L=1$ bound state, because the four $J=1/2$ orbitals are occupied by the four neutrons in the $\alpha$ clusters. Ignoring the spin degree of freedom, we simply assume the neutron in $^9$Be to occupy one of the lowest $L=1$ bound states in the square well modeling the $\alpha\alpha$ cluster core. The wave functions for the $L=0$ state in a spherical square well of radius $R_0$ are given by:
\ba
j_0(kr)Y_{00}(\Omega) &= \displaystyle\frac{\sin kr}{\sqrt{4\pi} kr}\,, \qquad & {\rm for}~r<R_0\,,
\nn \\[0.3cm]
h_0^{(1)}(i\kappa r)Y_{00}(\Omega) &= \displaystyle\frac{e^{-\kappa r}}{\sqrt{4\pi} \kappa r}\,, \qquad &{\rm for}~r>R_0\,.
\ea
The wave functions for the $L=1$ state are, respectively:
\ba
 &j_1(kr)Y_{10}(\Omega) = \displaystyle\frac{\sin kr - kr \cos kr}{(kr)^2} \sqrt{\frac{3}{4\pi}}\cos\theta \nn\\ & {\rm for}~r<R_0\,,\nn\\[0.4cm] 
&h_1^{(1)}(i\kappa r)Y_{10}(\Omega) = \displaystyle\frac{e^{-\kappa r}}{(\kappa r)^2}(\kappa r+1) \sqrt{\frac{3}{4\pi}}\cos\theta \nn\\ & {\rm for}~r>R_0 ;
\ea
At the edge of the square well, the radial wave functions and their radial derivatives must be continuous. This fixes the relative pre-factors and yields the eigenvalue equation for the binding energy:
\ba
\frac{k\cos kR_0}{\sin kR_0} =& -\kappa &\qquad (L=0)\,,\\[0.3cm]
\frac{(kR_0)^2\sin kR_0}{\sin kR_0 - kR_0 \cos kR_0} =&-\displaystyle\frac{(\kappa R_0)^2}{\kappa R_0 +1} &\qquad (L=1)\,.
\ea

An on-axis cut of the molecular neutron wave function for a symmetric superposition of the deuteron-centered bound state ($s$-wave, at +10\,fm) and the $^9$Be-centered bound state ($p$-wave, at $-10$\,fm) is shown in \rf{fig:Be9p_wavefunction}. In \rf{fig:Be9p_wavefunction3D} we show a 3-D representation of the same wave function.
\begin{figure} 
\centering
\includegraphics[width=0.8\columnwidth]{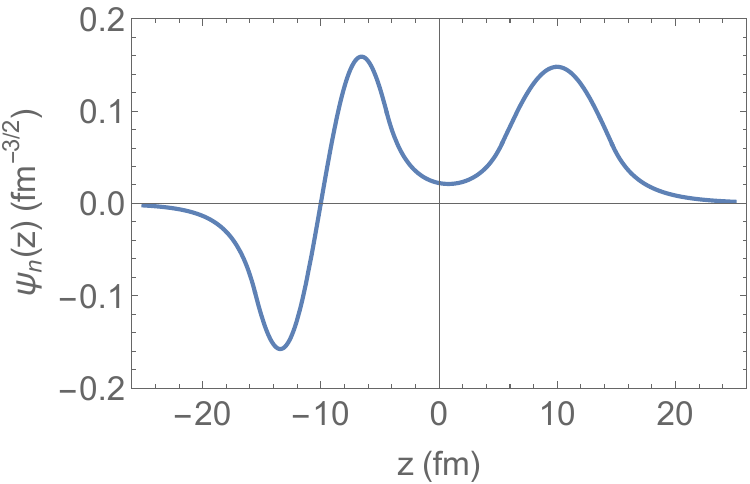}
\caption{On-axis cut of the shared neutron wave function in a symmetric linear superposition between $^9$Be (left) and the deuteron (right) for a $^8$Be and $p$ centers located at $\pm10$\,fm (separation of $R=20$\,fm). }
\label{fig:Be9p_wavefunction}
\end{figure}

\begin{figure} 
\centering
 \includegraphics[width= 0.8\columnwidth]{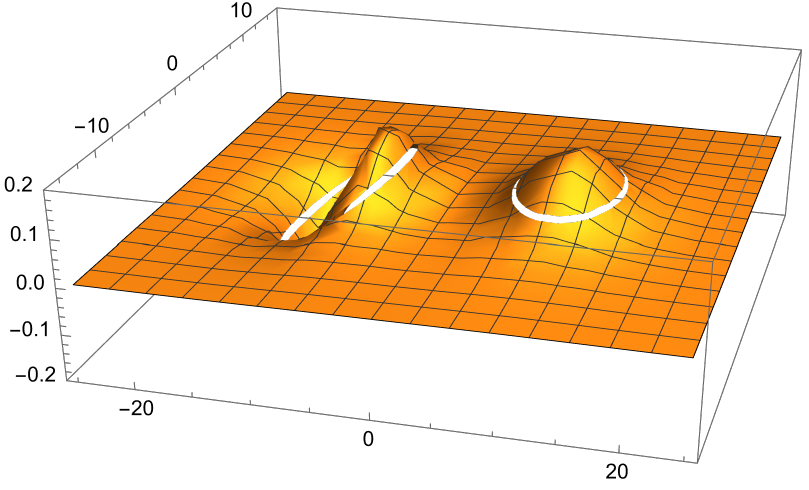}
\caption{3D representation of the shared neutron wave function in a symmetric linear superposition between $^9$Be (left) and the deuteron (right) for a $^8$Be and $p$ centers located at $\pm10$\,fm (separation of $R=20$\,fm). [The white lines are plot artefacts.]}
\label{fig:Be9p_wavefunction3D}
\end{figure}

When the proton and the $^8$Be core are separated by a distance $R$ much greater than the sum of the two square well radii, with the neutron being bound in both potentials, there is a dynamic tunneling barrier between the two wells that inhibits the transfer of the neutron from the $^8$Be core to the proton. The tunneling probability can be estimated from the instanton that exists inside the inverted ($V\to -V$) tunneling barrier. In order to estimate the action associated with this instanton we now calculate the energy splitting between two lowest energy eigenstates in the double-well potential. When $R\to\infty$, the energy eigenvalues are those associated with the neutron bound, either by the proton or by the $^8$Be core, and therefore the energy difference equals the asymptotic $Q$-value of the reaction: $\Delta E \equiv Q = E_2 - E_1 = E_{\rm b}(^9{\rm Be}) - E_{\rm b}(d) = 0.5592$ MeV. 

In the linear combination of nuclear orbitals (LCNO) approximation, we describe the true eigenstates of the nuclear molecule as linear superposition of the two separate nuclear states: When the two potential wells approach each other, the neutron feels both potential wells, and in our approach its wave function becomes a superposition of the asymptotic states with a distance-dependent phase $\alpha(R)$
\ba
\psi_+ &= \cos\alpha(R)\, \psi_1 + \sin\alpha(R)\, \psi_2,\\[0.2cm]
\psi_- &= -\sin\alpha(R)\, \psi_1 + \cos\alpha(R)\, \psi_2 ,
\label{eq:LCNO}
\ea
where the subscripts `1' and `2' refer to the neutron wave functions centered around the proton and the $^8$Be core, respectively. 

Similarly denoting the two potentials as $V_1$\,($p$-cenered) and $V_2$\,($^8$Be-centered), the Hamiltonian in the reduced basis spanned by the two ground states is
\be
\left(
\begin{array}{cc}
E_1 + \langle\psi_1 |V_2| \psi_1\rangle & E_2\langle\psi_1|\psi_2\rangle + \langle\psi_1 |V_1| \psi_2\rangle \\ 
 & \\
 E_1\langle\psi_2|\psi_1\rangle + \langle\psi_2 |V_2| \psi_1\rangle & E_2 + \langle\psi_2 |V_1| \psi_2\rangle
\end{array}
\right) 
\equiv\left(
\begin{array}{cc}
E'_1 & V_{12} \\ V_{21} & E'_2 
\end{array}
\right) ,
\ee
where $V_{12} = V_{21}^*$. The two energy eigenvalues are:
\be
E_\pm = \frac{E'_1+E'_2}{2} \mp \sqrt{ \frac{(E'_1-E'_2)^2}{4} + |V_{12}|^2 }.
\label{eq:E+-}
\ee
Any non-vanishing off-diagonal matrix element $V_{12}$ increases the splitting between the two states. 
 
\begin{figure}[htb]
\centering
\includegraphics[width=0.45\columnwidth]{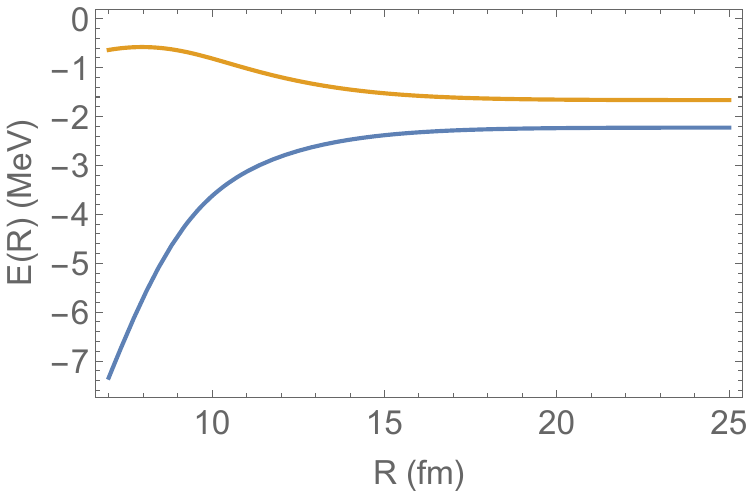}
\hspace{0.05\columnwidth}
\includegraphics[width=0.45\columnwidth] {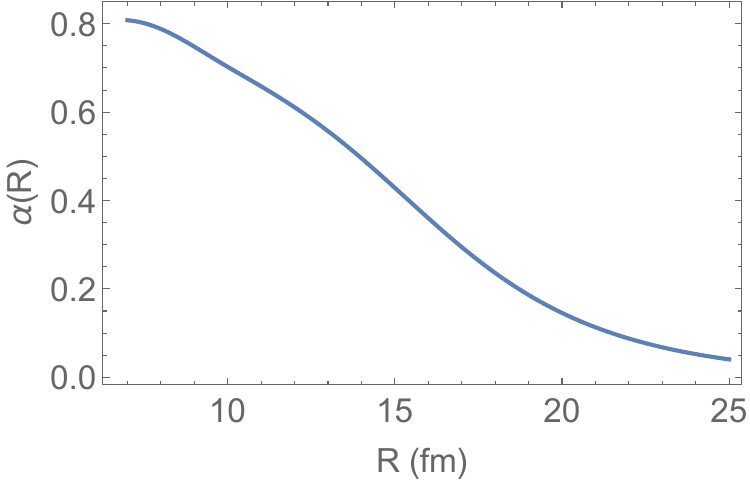}
\caption{Left panel: The blue and orange curves show the energy eigenvalues of the two orthogonal eigenvectors of the reduced Hamiltonian in units of MeV. The horizontal axis shows the $^8{\rm Be}-p$ distance $R$. The blue curve is the eigenstate that adiabatically connects with the deuteron ground state; the orange curve shows the state that adiabatically connects with the $^9$Be ground state. 
Right panel: Mixing angle $\alpha$ for the more deeply bound asymptotic $d$ state (blue curve in \rf{fig:Be9p_groundstates}) as a function of $R$.}
\label{fig:Be9p_groundstates}
\end{figure}

The energies of the two orthogonal eigenvectors of the reduced two-state Hamiltonian are shown in \rf{fig:Be9p_groundstates}. The more deeply bound blue curve is the eigenstate that connects for large $R$ with the deuteron ground state; the orange curve shows the orthogonal state that connects for large separation $R$ with the $^9$Be ground state. 

The right panel of \rf{fig:Be9p_groundstates} shows the mixing angle $\alpha(R)$ for the lower state (blue curve in the left panel) as function of $R$. At large $R$ the mixing angle approaches zero, and the state approaches the deuteron ground state.

\subsection{Sub-Barrier Neutron Transfer}

In order to describe the sub-barrier neutron transfer, we need to correctly describe the dynamics of the tunneling process. To do so, we write the full Hamiltonian of the $^8{\rm Be}-n-p$ system in two forms corresponding to the asymptotic configurations
\be
{\rm (I)}\quad [^8{\rm Be}+n]-p \equiv \, ^9{\rm Be}-p
\qquad
{\rm (II)}\quad ^8{\rm Be}-[n+p] \equiv \, ^8{\rm Be}-d .
\ee 
We also introduce the notations $P$ (CM momentum), $k_p$ (relative $p-~^9{\rm Be}$ momentum), $k_n$ (relative $n-~^8{\rm Be}$ momentum), $k_d$ (relative $d-~^8{\rm Be}$ momentum), and $k'_n$ (relative $n-p$ momentum) and the reduced masses
\be
{\rm (I)}\quad\mu_p = \displaystyle\frac{m_9 m_p}{M} \qquad
\mu_n = \frac{m_8 m_n}{m_9}
\qquad\qquad
{\rm (II)}\quad \mu_d = \displaystyle\frac{m_8 m_d}{M} \qquad
\mu'_n = \frac{m_p m_n}{m_d}
\ee
where  $m_8, m_9$ denote the masses of $^8{\rm Be}$ and $^9{\rm Be}$, respectively. $M$ is the total ten nucleon system mass -- considering the release of rest mass into  kinetic motion this is the CM-frame defined quantity and thus the same for the two configurations described below.

We can now write the full Hamiltonian in the following two equivalent forms:
\be
H \equiv H_{\rm I} = \frac{P^2}{2M} + \frac{k_p^2}{2\mu_p} + V_C(R) + \left[\left( \frac{k_n^2}{2\mu_n} + V_{8n} \right) + V_{pn} \right] ,
\ee
\be
H \equiv H_{\rm II} = \frac{P^2}{2M} + \frac{k_d^2}{2\mu_d} + V_C(R) + \left[\left( \frac{k_n'^2}{2\mu'_n} + V_{pn} \right) + V_{8n} \right] .
\ee
Here $V_{8n}, V_{pn}$ are the nuclear interactions of the neutron with the $^8$Be core and the proton, respectively, and $V_C(R) = 4e^2/R$ is the Coulomb potential between the proton and the $^8$Be core. We neglect the nuclear interaction between the proton and $^8$Be, because the neutron transfer occurs at distances $R$ where this interaction is negligible. We also note that the CM kinetic energy $P^2/2M$ is conserved and can be eliminated in both forms of $H$ by working in the CM system. 

As we noted earlier, the neutron in $^9$Be occupies a $p_{3/2}$ state, and the ground state wave function has negative parity. Because the $\alpha\alpha$ system has positive parity because of Bose symmetry, this implies that either in incident proton or the outgoing deuteron must have odd orbital angular momentum. At low energies this favors relative angular momentum $\ell=1$ in either the initial or the final state. We will denote these angular momenta as $\ell_p$ and $\ell_d$ with the constraint $\ell_p+\ell_d = 1$ with the associated centrifugal barriers $V_{L,i}(R) = \ell_i(\ell_i+1)/(2\mu_iR^2)$.

In the previous section we constructed approximate eigenstates of the part of $H$ enclosed by square brackets. After removing the CM term, the remainder of the Hamiltonian describes the relative motion, which is guided by the combined action of the Coulomb potential and the energies~\req{eq:E+-} associated with the bracketed part of the Hamiltonian. Because we are interested in very low relative motion energies we will treat its dynamics semi-classically. For the two states connecting with the asymptotic channels this gives:
\ba
{\rm (I)}\quad
(dR/dt)^2 &=& 2\mu_p \left[ E_{\rm cm} - V_{L,p}(R) - V_C(R) - E_{-}(R) \right] \equiv 2 \mu_p[E_{\rm cm}^{(p)}-U_{-}(R)]
\nn \\
{\rm (II)}\quad
(dR/dt)^2 &=& 2\mu_d \left[ E_{\rm cm} - V_{L,d}(R) - V_C(R) - E_{+}(R) \right] \equiv 2 \mu_d[E_{\rm cm}^{(p)}-U_{+}(R)].
\label{eq:rel-motion}
\ea
Here $E_{\rm cm}$ contains the $^9$Be binding energy, which is removed from $E_{\rm cm}^{(p)}$. The asymptotic values of the barrier potential are: $U_{-}(\infty) = 0$, $U_{+}(\infty) = E_{\rm b}(d)-E_{\rm b}(^9{\rm Be})$.
In the classically forbidden (tunneling) region the velocity is imaginary, corresponding to relative motion in imaginary time.

The dynamic barriers $U_\pm(R)$ defined in~\req{eq:rel-motion} are shown in \rf{fig:Be9p_Barrier} as a function of the internuclear separation $R$ out to 30~fm, not enough to fully reach their asymptotic values. Note that the energy scale has been normalized to correspond to the kinetic energy of the incident proton in the CM system. The solid blue (orange) curve represents the Coulomb and angular momentum barrier modified by the nuclear interaction for the more deeply (weakly) bound LCNO state that asymptotes to the $d+X$ ($^9{\rm Be}+p$) configuration, i.e.\ the blue (orange) curve in the left panel of \rf{fig:Be9p_groundstates}. The dashed lines show the Coulomb and anguar momentum barrier only. One can see that the molecular nuclear binding effect {\it i.e. shared orbitals} becomes relevant for internuclear distances below 15 fm.
\begin{figure} 
\centering
\includegraphics[width=0.45\columnwidth]{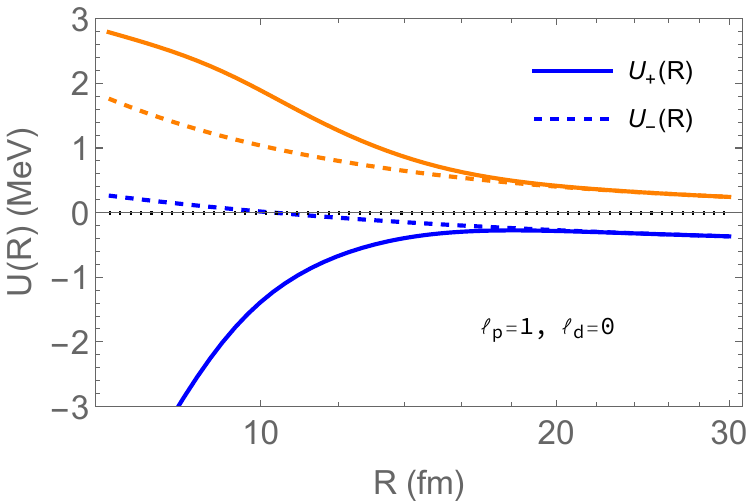}
\hspace{0.05\columnwidth}
\includegraphics[width=0.45\columnwidth]{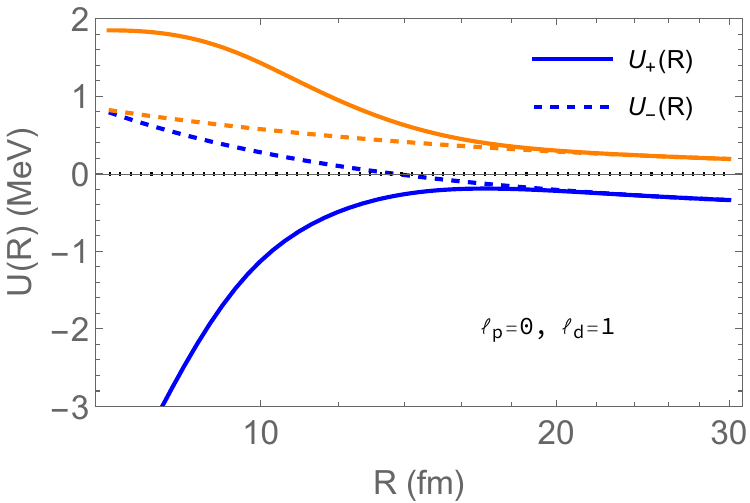}
\caption{The tunneling potentials $U_\pm(R)$ defined in~\req{eq:rel-motion} as a function of the internuclear separation $R$. Left panel: $\ell_p=1,\ell_d=0$, right panel: $\ell_p=0,\ell_d=1$. The solid orange curve shows the potential $U_{-}$ of the molecular state corresponding to the $^9{\rm Be}-p$ entrance channel; the solid blue curve depicts the potential $U_{+}$ of the molecular state corresponding to the $^8{\rm Be}-d$ exit channel. The dashed orange and blue curves show the respective Coulomb and angular momentum barriers without the nuclear interaction. The black dashed line indicates the $^9{\rm Be}-p$ threshold. The energy scale is normalized to the kinetic energy of the incident proton in the CM frame. The horizontal axis shows the nuclear separation $R$ in fm. The shape of the effective molecular potentials shown deserves explanation: Molecular energy curves for electronic molecules generally do not involve long-range Coulomb repulsion, because at least one partner is a neutral atom. This allows electron delocalization facilitated by the overlap between atomic orbitals to form a minimum in the molecular potential at rather large interatomic distances. In the nuclear case studied here, there exists a long-range repulsive force between the reaction partners, which is only overwhelmed by the nuclear orbital mixing effect at short distances $R < 15$ fm for the symmetric molecular state $\psi_+(R)$ shown as solid blue line in the figure.}
\label{fig:Be9p_Barrier}
\end{figure}
It is evident from \rf{fig:Be9p_Barrier} that for low proton energies $E_p \ll 1$ MeV the nuclear reaction in the entrance channel $^9{\rm Be}+p$ (solid orange line) always occurs in the classically forbidden (tunneling) region. However, once the neutron transfer has occurred, the energy is always above the barrier (solid blue line), and the system can easily proceed to fusion into $^{10}{\rm B}$ or to the $d+X$ exit channel. This conclusion is consistent with the observation that the partial widths for the subbarrier resonance at 319 keV into the neutron transfer channel $^9{\rm Be}(p,d)^8{\rm Be}$ and the compound nuclear channel $^9{\rm Be}(p,\alpha)^6{\rm Li}$ are of similar magnitude (see Fig.~6 and Table 3 in \cite{Zahnow:1997low}).

Next we inspect the neutron transfer probability. We treat the scattering as a time-dependent problem with the internuclear distance $R(t)$ a function of time. We expand the time-dependent wave function in the nuclear molecular eigenfunctions~\req{eq:LCNO} :
\ba
\psi(t) &=& \tilde{a}(t)\,\psi_+(R(t)) + \tilde{b}(t)\,\psi_-(R(t)) 
\nn\\
&=& [\tilde{a}(t)\cos\alpha(R) - \tilde{b}(t)\sin\alpha(R)]\, \psi_{\rm d} + [\tilde{a}(t)\sin\alpha(R) + \tilde{b}(t)\cos\alpha(R)]\, \psi_{\rm Be} .
\ea
The initial condition is $|\tilde{a}(-\infty)|=0, |\tilde{b}(-\infty)| = 1$.The coupled differential equations for the amplitudes $\tilde{a}(t)$ and $\tilde{b}(t)$ can be simplified by the phase transformation
\be
\tilde{a}(t) = a(t) e^{-i\chi_+(t)}, 
\qquad
\tilde{b}(t) = b(t) e^{-i\chi_-(t)}
\ee
with the adiabatic phases 
\be
\chi_\pm(t) = \int_{-\infty}^{t} dt'\, E_\pm(R(t')) .
\ee
The equations for $a(t)$ and $b(t)$ then take the form
\ba 
\frac{da}{dt} = - b(t) \langle\psi_+ | \partial\psi_-/\partial t \rangle\, e^{i(\chi_+ - \chi_-)}\,,
\qquad 
\frac{db}{dt} = - a(t) \langle\psi_- | \partial\psi_+/\partial t \rangle\, e^{i(\chi_- - \chi_+)} .
\label{eq:cceq}
\ea 
The final probabilities for finding the neutron in the deuteron and $^9$Be are, respectively:
\be
P({\rm d}) = |a(\infty)|^2, \qquad
P(^9{\rm Be}) = |b(\infty)|^2 .
\ee

Using the expansion in terms of the asymptotic nuclear wave functions the matrix elements take the simple form:
\ba
\langle\psi_- |\partial\psi_+/\partial t\rangle 
= -\langle\psi_+ |\partial\psi_-/\partial t\rangle
 = \frac{d\alpha(R(t))}{dt}\,,
\qquad 
\langle\psi_+ |\partial\psi_+/\partial t\rangle 
= \langle\psi_- |\partial\psi_-/\partial t\rangle 
 = 0 .
\ea
With this result the equations (\ref{eq:cceq}) simplify:and take the form
\ba
\frac{da}{dt} = +b(t)\,\displaystyle\frac{d\alpha}{dt}\, e^{i(\chi_+ - \chi_-)}\;,
\qquad 
\frac{db}{dt} = - a(t)\,\displaystyle\frac{d\alpha}{dt}\, e^{i(\chi_- - \chi_+)}\; ,
\ea
which are the general Landau-Zener equations for an avoided level crossing. It is easy to verify that the equations conserve the total probability $|a(t)|^2+|b(t)|^2 =1$.

If the time-dependence of the $^9$Be-$p$ distance is known, the equations can be written in the form
\ba
\frac{da}{dR} = +b(R)\,\displaystyle\frac{d\alpha}{dR}\, e^{i(\chi_+ - \chi_-)}\,,
\qquad 
\frac{db}{dR} = - a(R)\,\displaystyle\frac{d\alpha}{dR}\, e^{i(\chi_- - \chi_+)} 
\ea
with 
\be
\chi_\pm(R) = \int_{\infty}^{R} dR'\, \frac{E_\pm(R')}{dR'/dt} .
\ee
In the classically forbidden region, we can analytically continue $R(t)\to R(-i\tau)$, which means that
\be
\chi_\pm(R) \to -i \int dR'\,\frac{E_\pm(R')}{dR'/d\tau} \equiv -i{\tilde\chi}_\pm(R),
\ee
yielding an exponential damping factor $e^{-\tilde\chi(R)}$ instead of an oscillating phase factor. 

Finally, we give the expression for the molecular fusion cross section, noting that only a single partial wave ($L=1,m=0$) contributes to the molecular enhanced reaction:
\be
\sigma^{(\rm mol)}_{^9{\rm Be}(p,d)^8{\rm Be}}(E_p) = \frac{2\pi}{\mu_p E_p} |a(\infty,E_p)|^2 .
\ee
The strong enhancement of the instanton driven reaction rate requires a specific geometric configuration of the two colliding reactants.  This translates into a less dramatic enhancement of the traditional low energy nuclear $S$-factor comprising all possible collision geometries. 

\section{Summary, Context, and Outlook}

We have described the unique potential of $^9$Be in aneutronic fusion. We followed up proposing a novel nucelar reaction model suitable for systems undergoing neutron transfer at a relatively large distance, that is at low collision energy. We note that low collision energy reactions can lead to relatively small separation of nuclei for example due to electron  screening of the Coulomb $p-^9$Be Coulomb repulsion by high -$Z$ catalysts~\cite{Grayson:2025kva}.

Our semiclassical Born-Oppenheimer molecular model treatment considers the collision system $^9\mathrm{Be} + p$ as a nuclear quasi-molecule where the neutron is asymptotically bound to either a $^8$Be=$\alpha+\alpha$ core or to the proton. The quasi-molecular wave function of the neutron is expressed as the linear combination of these two asymptotic states of the neutron. In our model, the neutron transfer reaction at low energy is obtained as a resonant ``instanton'' tunneling process between the two asymptotic configurations. The transfer probability is dominated by sub-barrier transitions between the two lowest quasi-molecular states.

Our model can be refined by including the neutron's spin-orbit interaction, which lowers the p$_{3/2}$ state below the p$_{1/2}$ state in $^9$Be, a realistic radial dependence of the nuclear force, and a more accurate treatment of the reduced mass of the neutron outside the asymptotic regions. A more sophisticated model based on the resonating group method would treat the transfer reaction as a four-body problem ($\alpha, \alpha, p, n$) with two asymptotic channels ($p + ^9$Be and $\alpha+\alpha$ + d).

The physical environment for fusion we have in mind differs profoundly from the thermal burn of deuterium and tritium $d+t$ producing high energy free neutrons. We seek multi-step reaction cycles leading to aneutronic nuclear energy generation. Such  approach to fusion energy may not have received the appropriate prior attention. Our idea is supported by the observation that the solar core nuclear fusion ``reactor'' operates on two  cycles, the  $pp$- and CNO-reaction cycles. Both multi-step cycles are completely aneutronic - to the best of our knowledge and understanding there are no free neutrons produced. One must assume that there are many other aneutronic cycles which await discovery and exploration. Another advantage of such an approach to nuclear fusion is that in general one can expect cycles to exist which operate entirely on materials that can be found in natural mineral sources .

Our primary scientific interest is furthermore focused on non-equilibrium fusion environments created in interactions of ultra-short, high-contrast laser pulses~\cite{Labaune:2013dla}, with nano-structured reactants considered more recently~\cite{Biro:2023,Biro:2025,Csernai:2025,Kroo:2025}. In this situation no steady state conditions arise. In this aspect we advocate for nano-sized fusion unlike the large majority of the present day plasma physics attempts to harness fusion energy, such as ITER (magnetic confinement) or NIF (inertial confinement), which in comparison are very large scale plasma burn systems. For the successful implementation of those approaches the ongoing research problems are within engineering challenges not requiring improved understanding of the nuclear physics aspects of the reactions involved.

We hope that this short contribution provides compelling motivation to return to the scientific root of fusion: nuclear science. We need to study and investigate aneutronic nuclear reaction cycles not accessible to inertial and magnetic confinement plasma systems. The reader should view the present work as a first step along a long road leading to a more comprehensive understanding of low energy nuclear transmutation aneutronic fusion reactions and the associated development of novel nuclear reaction cycles.\\

\textbf{Acknowledgment:} We thank Tamas Bir\'o and the Wigner Hun-Ren Research Center for their kind hospitality in Budapest during the PP2024 conference, supported by NKFIH (Hungarian National Office for Research, Development and Innovation) under awards 2022-2.1.1-NL-2022-00002 and 2020-2.1.1-ED-2024-00314. This meeting and the related research report motivated the presentation of these recently obtained results.\\
\indent\textbf{Funding Statement:} Work of JR is not supported by any sponsor. BM acknowledges support from the U.S. Department of Energy, Office of Science, Nuclear Physics (grant DE-FG02-05ER41671).\\
\indent\textbf{Data Availability:} No datasets were generated or analyzed during the current study.\\
\indent\textbf{Competing Interests:} The authors declare no competing interests.


\begin{thebibliography}{9}

\bibitem{Moschini:2021}
L. Moschini, N. K. Timofeyuk
and R. C. Johnson,
``Perturbative correction to the adiabatic
approximation for $(d, p)$ reactions,''
J. Phys. G: Nucl. Part. Phys. 48 (2021) 095102 

\bibitem{Kelley:2012} 
J.H. Kelley, E. Kwan, J.E. Purcell, C.G. Sheu and H.R. Weller
``Energy levels of light nuclei A = 11,''
Nuc. Phys. A \textbf{880} (2012) 88–195

\bibitem{Koll:2025} Koll, D., Lachner, J., Beutner, S. et al.,
``A cosmogenic $^{10}$Be anomaly during the late Miocene as independent time marker for marine archives,'' 
Nat Commun \textbf{16} (2025) 866 

\bibitem{Le_Deunff_2010}
J. Le Deunff and A. Mouchet,
``Instantons re-examined: Dynamical tunneling and resonant tunneling,''
Phys. Rev. E \textbf{81} (2010) 046205
[arXiv:0911.4093 [quant-ph]].

\bibitem{Barker:2001uvh}
F.~C.~Barker and Y.~Kond\={o},
``The 9Be (p,$\alpha$) 6Li and 9Be $(p,d)$ 8Be cross sections at low energies,''
Nucl. Phys. A \textbf{688} (2001) 959-974

\bibitem{Macfarlane:1960zz}
M.~H.~Macfarlane and J.~B.~French,
``Stripping Reactions and the Structure of Light and Intermediate Nuclei,''
Rev. Mod. Phys. \textbf{32} (1960) 567-691

\bibitem{Rakityansky:2024zew}
S.~A.~Rakityansky,
``Wave function of Be9 in the three-body (\ensuremath{\alpha}\ensuremath{\alpha}n) model,''
Phys. Rev. C \textbf{110} (2024) 024001
[arXiv:2407.17371 [nucl-th]].

\bibitem{Nortershauser:2008vp}
W.~Nortershauser, \textit{et al.}
``Nuclear Charge Radii of Be-7, Be-9, Be-10 and the one-neutron halo nucleus Be-11,''
Phys. Rev. Lett. \textbf{102} (2009) 062503
[arXiv:0809.2607 [nucl-ex]].

\bibitem{Machleidt:1987hj}
R.~Machleidt, K.~Holinde and C.~Elster,
``The Bonn Meson Exchange Model for the Nucleon Nucleon Interaction,''
Phys. Rept. \textbf{149} (1987) 1-89 

\bibitem{Zahnow:1997low}
D. Zahnow, C. Rolfs, S. Schmidt, and H. P. Trautvetter, 
``Low-energy S(E) factor of
$^9{\rm Be}(p,\alpha)^6{\rm Li}$
and $^9{\rm Be}(p,d)^8{\rm Be}$,''
Z. Phys. A: Hadrons and Nuclei, \textbf{359} (1997) 211 

\bibitem{Labaune:2013dla}
C.~Labaune, C.~Baccou, S.~Depierreux, C.~Goyon, G.~Loisel, V.~Yahia and J.~Rafelski,
``Fusion reactions initiated by laser-accelerated particle beams in a laser-produced plasma,''
Nature Commun. \textbf{4} (2013) 2506
[arXiv:1310.2002 [physics.plasm-ph]]

\bibitem{Grayson:2025kva}
C.~Grayson and J.~Rafelski,
``Nuclear Fusion Enhancement by Heavy Nuclear Catalysts,''
[arXiv:2502.07804 [physics.plasm-ph]] EPJSt (2025) in this volume 

\bibitem{Biro:2023}
T.S. Bir\'o, {\it et al.},
``With Nanoplasmonics Towards Fusion,''
Universe \textbf{9} (2023) 233 

\bibitem{Biro:2025}
T.S. Bir\'o, for the NAPLIFE collaboration,
``Nanotechnology and Plasmonics for Fusion,'' \href {https://doi.org/10.1140/epjs/s11734-025-01562-7}{Eur. Phys. J. Spec. Top. (2025).}, in this volume.

\bibitem{Csernai:2025}
L.P. Csernai, for the NAPLIFE Collaboration, ``High-energy non-thermal, laser-induced nano-fusion,''\href{https://doi.org/10.1140/epjs/s11734-025-01466-6a}
{Eur. Phys. J. Spec. Top. (2025).}, in this volume.

\bibitem{Kroo:2025}
N. Kroo, ``High field nanoplasmonics for nuclear fusion,'' Eur. Phys. J. Spec. Top. (2025). \href{https://doi.org/10.1140/epjs/s11734-025-01522-1}{Eur. Phys. J. Spec. Top. (2025).}, in this volume.
\end{thebibliography}
\end{document}